\documentclass[%
 aip,
 amsmath,amssymb,
 reprint,%
]{revtex4-1}

\usepackage{graphicx}
\usepackage{dcolumn}
\usepackage{bm}

\usepackage[utf8]{inputenc}
\usepackage[T1]{fontenc}
\usepackage{mathptmx}
\usepackage{etoolbox}
\usepackage{xcolor,soul}

\makeatletter
\def\@email#1#2{%
 \endgroup
 \patchcmd{\titleblock@produce}
  {\frontmatter@RRAPformat}
  {\frontmatter@RRAPformat{\produce@RRAP{*#1\href{mailto:#2}{#2}}}\frontmatter@RRAPformat}
  {}{}
}%
\makeatother
\begin{document}

\preprint{AIP/123-QED}

\title[Dielectric response in proteins: The proteotronics approach]{Dielectric response in proteins: The proteotronics approach}
\author{E. Alfinito}
\affiliation{ 
Dipartimento di Matematica e Fisica"Ennio De Giorgi", Università del Salento, Lecce, Italy 
}%
\email{eleonora.alfinito@unisalento.it}

\author{M. Beccaria}%
\affiliation{ Dipartimento di Matematica e Fisica"Ennio De Giorgi", Università del Salento, Lecce, Italy. }

\affiliation{%
National Institute for Nuclear Physics, INFN, Sezione di Lecce, Lecce, Italy
}%

\date{\today}

\begin{abstract}
The dielectric properties of proteins, particularly in their hydrated state, have been extensively studied. Numerous theoretical and experimental investigations have reported values of both the permittivity and the intrinsic dipole moments of specific proteins under well-defined hydration conditions. Since even approximate estimates of these properties are relevant from both fundamental and applied perspectives, we propose a {ease-to-use} method to calculate the relative permittivity that can be readily integrated into proteotronics workflows. To validate the proposed approach, we compare the results with those obtained using a classical macroscopic method. The outcomes are consistent and contribute further insight into this long-debated issue.

\end{abstract}

\maketitle

\section{\label{sec:intro} Introduction}

The dielectric response of bulk materials under small applied fields and steady-state conditions is well described within the framework of classical electrodynamics. Classical formulations account for both electronic and intrinsic (orientational) polarization and are independent of the material’s shape and size. In contrast, in nanostructured materials the dielectric response is strongly influenced by geometrical factors, including shape and size \cite{mo1995characteristics, riande2004electrical}. This aspect is particularly relevant for proteins, \textit{i.e.}, macromolecules characterized by highly irregular, non-geometric shapes, whose structure and functionality are stabilized in the hydrated state \cite{bellissent2016water}.

Proteins are composed of amino acids arranged in a linear sequence referred to as the primary structure. Through the folding process, this sequence adopts a specific three-dimensional conformation corresponding to the native state (tertiary structure). This native conformation is conserved among proteins of the same type and, more broadly, within the same family \cite{donnelly1994seven}. The solvent environment plays a crucial role in both the formation and stabilization of the native structure. In particular, water molecules located at the protein surface and within internal cavities provide electrostatic screening of the electric fields generated by charged and polar residues. Their ability to dynamically reorient in response to these fields contributes significantly to conformational stability. In conditions of insufficient hydration, proteins may lose their functional native state and eventually undergo denaturation.
The overall dielectric response of a protein, including the orientation and fluctuations of its dipole moment, strongly depends on the surrounding solvent and, more specifically, on the degree of hydration \cite{simonson1996charge, simonson2003electrostatics, simonson2008dielectric, schutz2001dielectric, warshel1998, warshel2006modeling}.

At the beginning of the 20th century, some seminal papers \cite{onsager1936electric, kirkwood1936statistical, kirkwood1939dielectric, frolich1949theory} began to investigate the significance of permittivity in dense polar materials. These studies laid the foundation for analyzing the dielectric response of both polar and nonpolar substances in polar environments, initially through analytical approaches \cite{tanford1957location, tanford1957theory} and later using computational methods \cite{antosiewicz1995computation, havranek1999tanford}.
Most subsequent investigations of proteins in solution build upon and adapt the Tanford–Kirkwood framework\cite{tanford1957location, tanford1957theory}. This model distinguishes an internal region of the protein, effectively shielded from solvent effects, and an external region in direct contact with the solvent. The relative dimensions of these regions largely determine the protein’s effective dielectric constant \cite{simonson1996charge,simonson2003electrostatics, simonson2008dielectric}.

 These computational approaches, including molecular dynamics (MD) and Monte Carlo (MC) simulations, are known to be time-consuming {  \cite{matyushov2012dipolar, amin2020variations,hartl2022dipolar, tenenbaum2024energy, agelii2025dipole}}, and the results obtained do not always agree with the limited available experimental data . To address this complexity, coarse-grained methods have been developed more recently, based on a continuous distribution of dielectric permittivity within the protein \cite{li2013dielectric, grant2001smooth}. The central idea of this kind of approach {(smooth permittivity} \cite{li2013dielectric}) is to provide an effective description of the protein–solvent system: in surface regions and internal cavities the effective permittivity is taken as an average between water and the protein residues, while within the protein core it approaches the permittivity of the dry residues. A key challenge is determining how to generate this spatial distribution. In the literature, this is often achieved by representing the protein as an equivalent sphere and continuously distributing permittivity values among its amino acids or atoms, based on measures such as the radius of gyration \cite{ li2013dielectric}. However, this approach can yield inconclusive results for highly elongated proteins \cite{li2013dielectric}.

{Following the continuous method, we propose to evaluate the effective permittivity by evaluating the shape of the protein not by the radius of gyration, but rather by the number} of nearest neighbors (coordination number) for each amino acid. This quantity is a key descriptor of the topological features of a protein’s specific conformation.
Specifically, it is used to produce the protein representations through complex networks in a procedure is called proteotronics\cite{alfinito2008network,eleonora2015proteotronics}. Furthermore, it is commonly used to generate two-dimensional maps of the tertiary structure, such as contact maps \cite{vendruscolo1997recovery,alfinito2010single}. 

{ The coordination number allows to easily locate the amino acids inside the protein, i.e. the  lower this number, the more superficial is the amino acid. On the other side, the more internal is the amino acid , the less effect the solvent has on it. }
Based on this approach, permittivity values are assigned so that highly connected regions correspond to lower permittivity, while less connected regions are assigned higher values. This strategy provides a more accurate estimate of buried versus exposed regions, thereby addressing limitations reported in previous studies \cite{li2013dielectric}. 

To define nearest neighbors, we model the protein as a complex network in which nodes correspond to the centroids of amino acids \cite{eleonora2015proteotronics,alfinito2011human, alfinito2010single}. For each node, nearest neighbors are defined as those nodes located within a distance less than $R_C$, the interaction radius.	
The choice of $R_C$
  determines the number of nearest neighbors: very small values (< 4 \AA) produce a disconnected network, where most nodes have only 0–1 neighbors, whereas very large values result in a fully connected network, with each amino acid linked to all others \cite{alfinito2008network,eleonora2015proteotronics, alfinito2011human}. Neither of these extremes is of interest. In this study, we set $R_C=6$ \AA \ which captures the geometric regularity of proteins \cite{simonson2003electrostatics,alfinito2008network,alfinito2010single} and yields an average of approximately five nearest neighbors per node.

Finally, an appropriate distribution function must be selected. In principle, different choices may lead to significantly different outcomes. To address this issue, we evaluated several candidate functions, seeking those that yield higher permittivity values for elongated proteins than for spherical ones, while ensuring consistency across structurally similar proteins. These functions are formulated using nearest-neighbor descriptors, including the maximum value  ($L_{\rm max}$) and 
the mean value ($\langle L\rangle$) computed over the entire protein. 
{Overall, only minor differences were observed among the test-functions, particularly when normalized data were considered.}
 
{Finally, we compare our data with those obtained from a macroscopic procedure that involves calculating the permittivity of the protein's electric dipole . This type of investigation is allowed by the Protein Dipole Moment Server} \cite{felder2007server}{developed in 2007 by Felder, Prilusky, and Sussman, which calculates the dipole moment of most protein entries in the protein data bank (PDB)}\cite{berman2000protein}.{ Remarkably, the agreement is  quite good and  suggests further investigations .}

\begin{figure*}
    \includegraphics[scale=0.50]{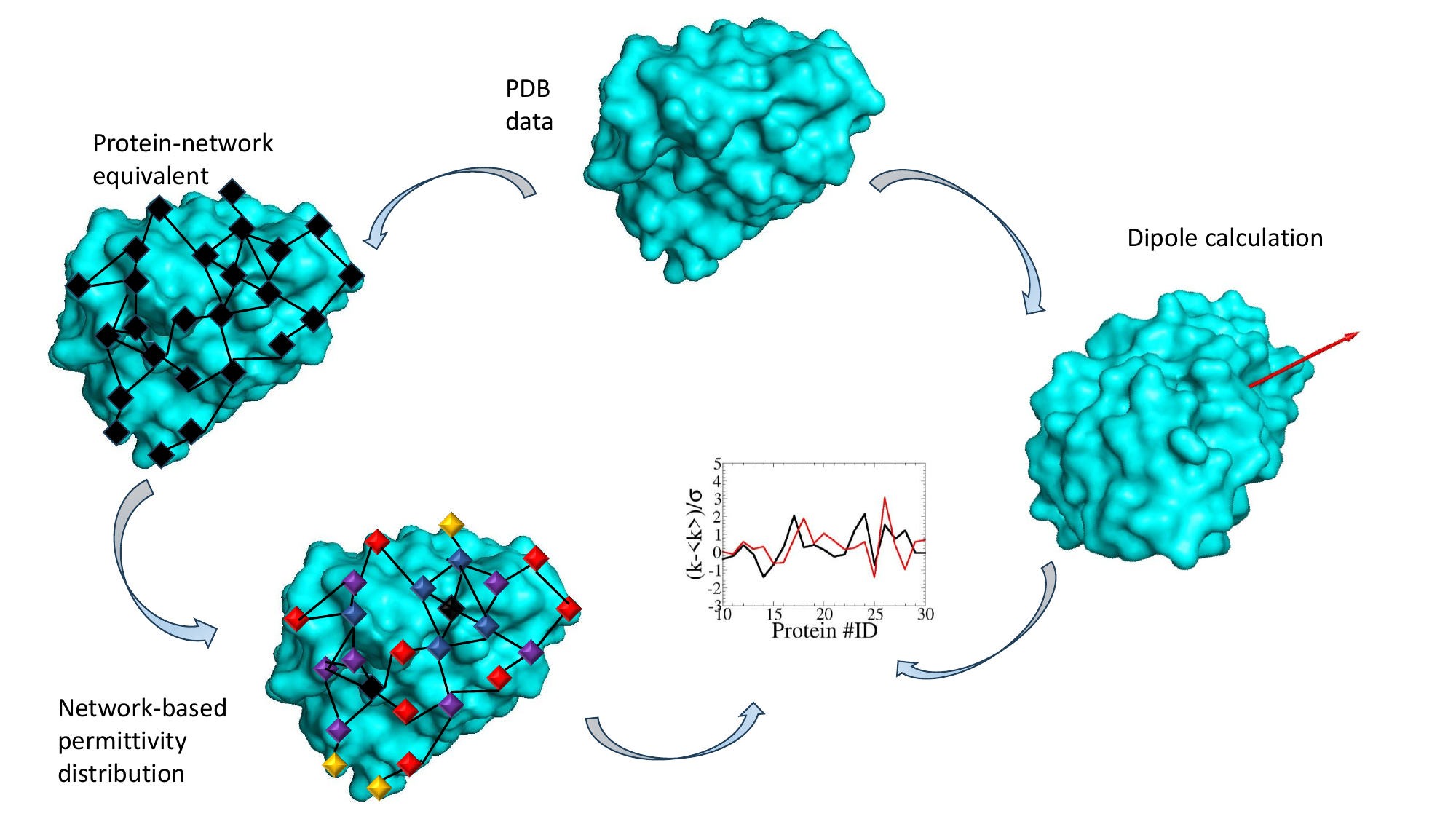}
    \caption{ {Summary of the research described in this article. a. The 3D structure of the protein is obtained from the Protein Data Bank; b. As part of the proteotronic approach, the structure is mapped onto a complex network; c. The permittivity is calculated from the coordination number; d. The protein dipole, calculated as in} Ref.\onlinecite{felder2007server}, {is used to produce an alternative value of the protein permittivity; e. Finally, the data obtained from the two different approaches are compared. To make the figure we
    computed charges by the tool PDB2PQR (https://www.poissonboltzmann.org) with AMBER force field, and visualised the corresponding dipole momentum inside
    PyMol ( https://www.pymol.org ). } }
    \label{fig:g_abstract}
\end{figure*}

{In Fig}\ref{fig:g_abstract} {a schematic view of the entire procedure is resumed.}

\section{\label{sec:sec2}Model and Results}
{In this section, we demonstrate two different procedures: one microscopic, inspired by continuum models, and the other macroscopic, based on a classical electrodynamics approach.

The microscopic approach is implemented using some test functions, as described in detail in subsection A, while the macroscopic approach, subsection B, demonstrates how to distinguish the dielectric response of a hydrated and dry protein. Finally, we compare the results obtained with the two methods.}
\subsection{\label{sec:level1} Continuous model: Test-functions}
{We consider a pool of test-functions  employed to assign a local permittivity value, 
$\epsilon(n)$, to each $n$-th amino acid.}\footnote{Hereafter, the term permittivity refers to the relative permittivity, i.e., the permittivity normalized to the vacuum permittivity.} This quantity depends on both the intrinsic (dry) permittivity, 
$\epsilon_{i}(n)$, and the amino acid’s position within the native protein structure. More specifically, $\epsilon(n)$
 varies between a maximum value corresponding to that of the solvent under standard temperature and pressure (STP) conditions and  $\epsilon_i(n)$, obtained using the polarizability values calculated in \cite{song2002inhomogeneous}. More details are given in Appendix \ref{app:appB}.
 We use three quite standard test-functions, namely, linear, sigmoidal and exponential,  whose 
sharpness can be tuned via a control parameter (Appendix \ref{app:appA}). Increasing the sharpness enhances the contrast between regions characterized by high and low permittivity, facilitating the identification of hydrated (“wet”) regions associated with larger 
$\epsilon(n)$ values.
The distinction between an internal low-permittivity region and an external high-permittivity region underlies many theoretical and computational studies \cite{simonson1996charge, simonson2003electrostatics, simonson2008dielectric, warshel2006modeling}, as the effective dielectric response critically depends on this partitioning.

Finally, the equivalent permittivity of the protein is defined as
$k=\langle \epsilon(n)\rangle$,
i.e., the average of the local permittivity values over all amino acids. This quantity depends on the specific choice of the test function and on its sharpness. Consequently, the control parameters can, in principle, be adjusted to reproduce available experimental data.
More importantly, however, the proposed procedure should capture the expected trends in $k$. In particular, $k$
is expected to assume lower values for compact, approximately spherical proteins and higher values for elongated structures, while remaining comparable across proteins with similar conformations.

\begin{figure}
\includegraphics[scale=0.33]{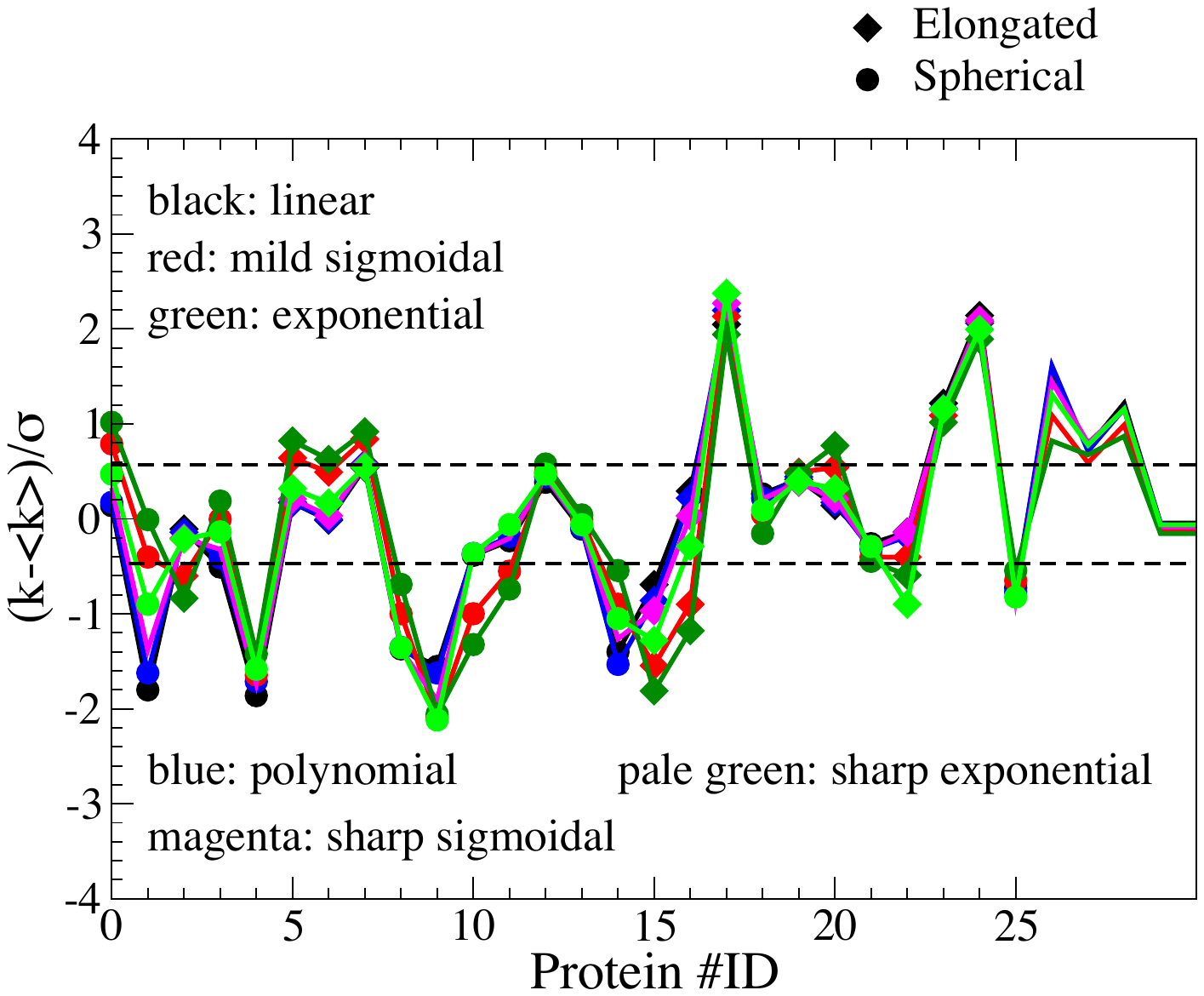}
\caption{ 
Normalized effective permittivity, $k$, for the selected dataset. The calculation is performed using three test functions (linear, sigmoidal, and exponential). Proteins with different geometrical characteristics (approximately spherical and elongated) are indicated by distinct symbols (dots and diamonds, respectively). Solid lines serve as guides to the eye. Dashed lines denote the mean values for spherical (lower) and elongated (upper) proteins. The complete protein dataset is listed in Table~\ref{tab:table1}.
}
\label{fig:fig.tutti}
\end{figure}

For this reason, in this paper we analyze a dataset consisting of 31 proteins, including both single-chain and complex structures, { elongated and spherical} structures, selected to span a range of shapes and sizes and to perform different functions.  The aim is to assess whether the proposed approach satisfies the minimal consistency requirements discussed above.{ The selected proteins constitute an unbiased sample and could be expanded in future studies.}
{To perform a consistent comparison among different models,} Figure \ref{fig:fig.tutti} presents the normalized permittivity values, $(k - \langle k \rangle)/\sigma$, where $\langle k \rangle$ denotes the mean and $\sigma$ the standard deviation computed over the dataset. The set of test functions considered
captures the differences between elongated and approximately spherical proteins. In particular, elongated proteins systematically exhibit  values of $k$ (see Figure \ref{fig:fig.tutti} ). { 
On the other hand, these differences are generally not so large as to be considered a weakness of the method, as in different papers\cite{li2013dielectric}. Specifically, for the pair of elongated 1uz3 and spherical 1bkr proteins, we calculate, with the proposed model functions, for the elongated protein, values of $k$ about 15-20$\%$ larger, while in that paper, the it has a permittivity value about 8 times larger, because the use of the radius of gyration leads to the inclusion of too much water\cite{li2013dielectric}.}

All test-functions show a comparable trend, which suggests that it is not the chosen function itself that is important, but rather the permittivity distribution based on the coordination number. For illustration, Fig. \ref{fig:fig.tutti} reports the dataset-averaged permittivity obtained using the linear distribution function, shown separately for elongated and spherical proteins.
Further details on the model functions are provided in Appendix A.

Finally, we compare previous results with those obtained using a macroscopic approach based on the evaluation of the protein’s electric dipole moment. This analysis is supported by the Protein Dipole Moment Server \cite{felder2007server}, developed by Felder, Prilusky, and Sussman, which provides dipole moment estimates for a large number of protein structures available in the Protein Data Bank (PDB) \cite{berman2000protein}. 
\subsection{\label{sec:level2} Macroscopic susceptibility}

To recall, within the framework of classical electrodynamics, the electrical polarizability of a collection of weakly interacting dipoles is determined by the ensemble-averaged dipole orientation induced by an external electric field, $E$. {
Each dipole rotates to reduce the potential energy, say $U=-\boldsymbol{p_0}\cdot\boldsymbol{E}$. On the other hand, thermal fluctuations work against perfect alignment, such that for a freely rotating dipole, the higher the temperature, the less likely it is to remain antiparallel to the applied field $\boldsymbol{E}$.}

The derivation follows from the thermal angular distribution function describing dipole orientations
\begin{equation}
\mathcal{P}(T)\,d\cos\theta=\frac{e^{-U/k_BT}dU}{\int{e^{-U/k_BT}dU}}, \;\; U=-\boldsymbol{p_0}\cdot\boldsymbol{E}=-p_0\,E\,\cos\theta,
 \label{eq:eqprob}
\end{equation}
where $k_B$ is the Boltzmann constant, and $\boldsymbol{p_0}$ the dipole momentum vector.
The mean projection of dipole momentum is 
\begin{equation}
    \langle p^\parallel(T)\rangle =\int{p_0\,\cos\theta\,\mathcal{P}}(T)d\cos\theta,
    \label{eq:eq2dipole}
\end{equation}
where $p^{\parallel} = p_0 \cos\theta$ denotes the component of the intrinsic dipole moment parallel to the applied electric field, i.e., the only component contributing to field screening. In the weak-field limit, the polarizability describes the macroscopic response of the dipole ensemble to the applied field. It is defined in terms of the mean dipole orientation per unit volume:
\begin{equation}
\label{eq:eq3dipole}
\Pi_0=n\langle p^\parallel(T)\rangle=n\frac{E\,p_0^2}{3  k_B T} \equiv \epsilon_0\chi_0 E   ,
\end{equation}
where $n$ is the ensemble density.
 
The susceptibility due to intrinsic dipole orientation is thus
\begin{equation}
\label{eq:eq4dipole}
\chi_0=n\frac{p_0^2}{3\epsilon_0 k_B T}    .
\end{equation}

{Proteins generally possess a not-negligible electric dipole moment}\cite{simonson1996charge,de2000blue,simonson2003electrostatics}.

{ Functioning proteins are usually hydrated protein, since the }
protein native structure itself arises from hydration  through the folding process, which typically positions hydrophilic residues toward the solvent-exposed exterior and hydrophobic residues within the protein core. Water molecules near the surface dynamically reorient in response to local electrostatic fields, thereby reducing the internal electric interactions generated by charged and polar residues and contributing to structural stability. Consequently, the dipole moment reported in Ref. \onlinecite{felder2007server} reflects this solvent-mediated charge organization, {depends on single residue charges and dipoles  and on the global protein shape.} Finally, hydrated proteins cannot be regarded as freely rotating entities, as the surrounding solvent constrains their orientation in the presence of external electric fields. 
{ 
As a consequence, when considering an ensemble of hydrated proteins, the macroscopic response to an external electric  field, permittivity, is smaller than that produced by the same ensemble  when considered free of  rotating.}

In contrast, the dipole moment of the dry protein differs due to the absence of solvent-induced polarization and screening effects.

The dipole moment of a dry protein, denoted as $p_0$, can be estimated following the approach proposed by Song \cite{song2002inhomogeneous}. In that paper, the \textit{intrinsic} polarizabilities of amino acids in proteins were evaluated using molecular dynamics simulations. The term \textit{intrinsic} refers to a condition in which the dipole moment of each amino acid is assumed to be unaffected by external influences (e.g., solvent effects) and by interactions with other dipoles within the protein \cite{song2002inhomogeneous}.
This approximation corresponds to the limiting case of a dry protein, where the dipole moments of individual amino acids are sufficiently small to be considered nearly independent \cite{simonson2003electrostatics}, and solvent-induced polarization is absent. Within this framework, the relative dielectric constant of a protein can be evaluated as $k_0 = \langle \epsilon(n) \rangle$, where, for each amino acid, $\epsilon(n) = \epsilon_i$, i.e., the intrinsic dielectric constant obtained from Ref. \onlinecite{song2002inhomogeneous} (see Appendix \ref{app:appA}).
 The resulting value of $k_0$ is more or less the same across the selected proteins, say, $k_0 \approx 1.5$ , in fine agreement with the outcomes of  Ref. { {\onlinecite{simonson2003electrostatics,  seyedi2018dipolar} }(see also Appendix B for details).
 
{ As a final comment, a collection of dry proteins may be regarded as freely orientable in an external electric field, due to the absence of solvent-induced friction. 
Consequently, the dielectric response of the system,\textit{i.e.}, the dry susceptibility, $\chi_0$, can be described by} Eq.~\ref{eq:eq4dipole}.

{Otherwise, solvent both reduces the ability of  the dipole  of a hydrated protein  to orient in external electric fields and modifies its  value  from $p_0$ to $p$. In other terms, $p=p(p_0)$, which allows to calculate the expectation value of polarization under hydrated conditions as follows:}
\begin{equation}
\langle p^\parallel(T)\rangle=\int p\,\cos\theta\,\mathcal{P}(T)\,d\cos\theta.
\label{eq:eq5dipole}
\end{equation}
{ where $\mathcal{P}(T)$ has been introduced in Eq.} \ref{eq:eqprob}.
We therefore propose an \textit{effective} susceptibility,$\chi^H$, representing the maximal dielectric response of a hydrated protein (\textit{cf}. Eq.~\ref{eq:eq4dipole}):
\begin{equation}
\label{eq:eq6dipole}
\chi^H=\frac{p\,p_0}{3\epsilon_0\,\Omega\,k_B T}.
\end{equation}
Here, $n = 1/\Omega$, with $\Omega$ denoting the volume of the hydrated protein.
Expressing $p_0$ in terms of the dry susceptibility, $\chi_0$, (\textit{cfr.} \ Eq.\ref{eq:eq4dipole}) and the dry protein volume, $\Omega_0$, the effective susceptibility becomes:
\begin{equation}
\label{eq:eq7dipole}
\chi^H=\frac{p\sqrt{\chi_0\Omega_0}}{\sqrt{3\epsilon_0 k_B T}\,\Omega}.
\end{equation}

Finally, we consider the following assumptions: (1) experimental studies indicate no significant differences between the volumes of hydrated, $\Omega$, and dry, $\Omega_0$, proteins \cite{harpaz1994volume}; and (2) the dry susceptibility, $\chi_0=k_0-1$, is taken to be approximately constant across different proteins.
Under these approximations, Eq.~\ref{eq:eq7dipole} reduces to
\begin{equation}
\chi^H \approx \frac{p}{\sqrt{6 \epsilon_0 \Omega k_B T}},
\label{eq:eq8dipole}
\end{equation}
where $\chi_0 \approx 0.5$ has been assumed, as previously explained.

In the present analysis, the hydrated protein volume is approximated as $\Omega = N_{aa}\langle\Omega_{res}\rangle$, where $N_{aa}$ denotes the number of amino acids and $\langle\Omega_{res}\rangle$ is the mean residue volume. The latter is taken as 127 \text{\AA}$^{3}$ as in Ref. \onlinecite{amin2020variations}. This estimate is consistent with previous determinations reported in Ref.~\onlinecite{chen2015proteinvolume}.
As a final remark, Eq.~\ref{eq:eq8dipole} yields susceptibility values within the range typically reported for hydrated proteins \cite{simonson2003electrostatics}, \textit{cfr.} Tab.\ref{tab:table1}. In contrast, the standard expression,
\begin{equation}
\label{eq:eq9dipole}
\chi=\frac{p^2}{3\epsilon_0 \,\Omega\, k_B\, T},
\end{equation}
produces values that deviate significantly from this range, \textit{cfr.} Tab.\ref{tab:table1}.
Finally, we compare the equivalent macroscopic permittivity, $k^H = \chi^H + 1$ , say, $(p - p_0)$ model, obtained by  Eq.~\ref{eq:eq8dipole} together with dipole moment data from Ref.~\onlinecite{felder2007server}, with   permittivity obtained from the continuous microscopic models . Both data are normalized as in Fig.~\ref{fig:fig.tutti}. The comparison indicates qualitative consistency. In particular, Fig.~\ref{fig:fig2} shows the normalized permittivity calculated using the linear distribution function alongside the macroscopic estimate .
The agreement  is satisfactory enough, especially for larger proteins (ID $> 10$, corresponding to $N_{aa} > 130$). A further comparison using the standard macroscopic dielectric constant derived from Eq.~\ref{eq:eq9dipole} (the $(p - p)$ \ model) also shows comparable qualitative agreement.

\begin{figure}
\includegraphics[scale=0.33]{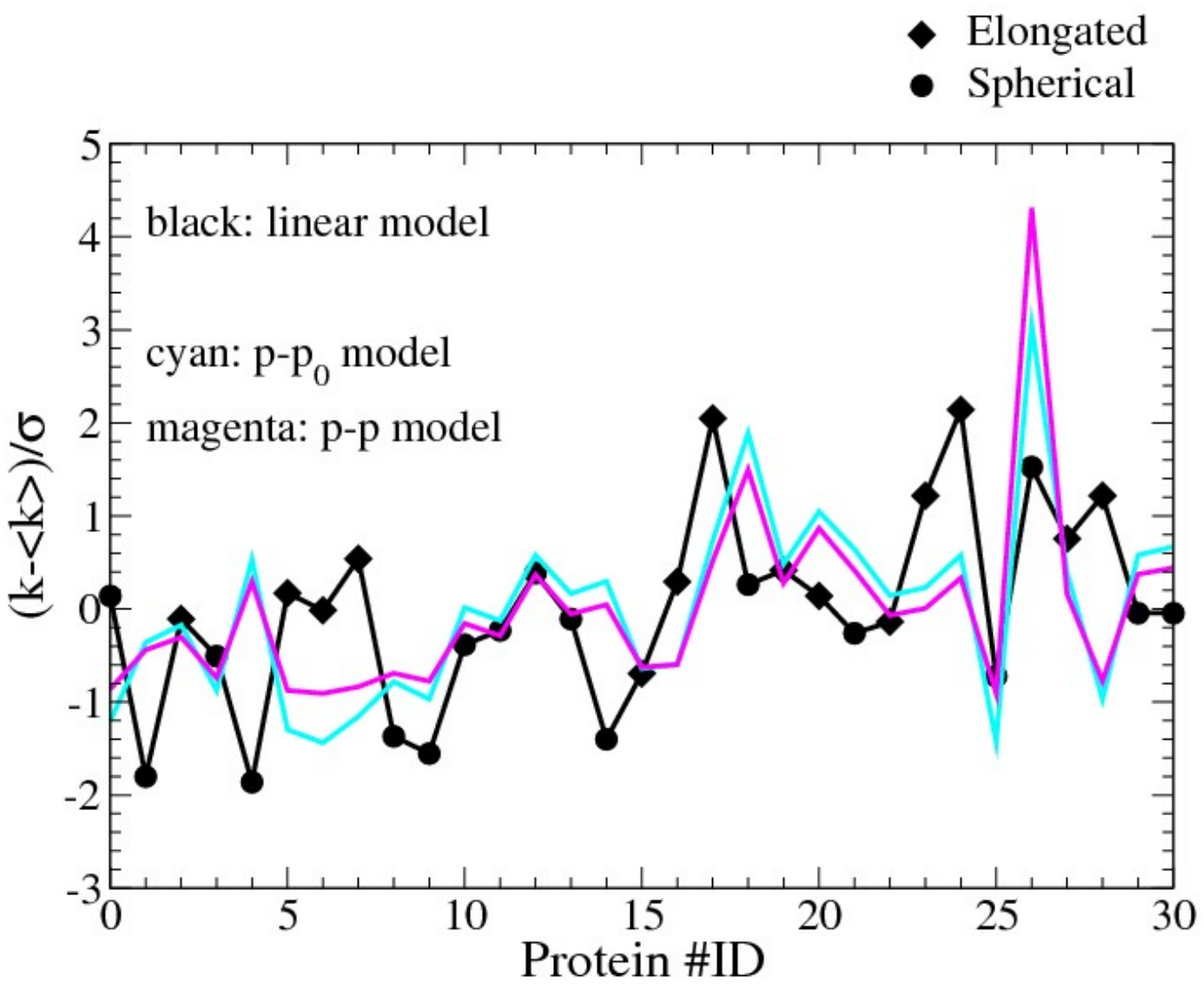}
\caption{\label{fig:fig2} 
Normalized \textit{macroscopic} susceptibility for the selected dataset: role of the solvent. The normalized permittivity values obtained using Eq.~\ref{eq:eq6dipole} (the $p - p_0$ model, cyan solid line) and Eq.~\ref{eq:eq9dipole} (the $p - p$ model, magenta line) are compared with those calculated using the linear test function (black line).
}

\end{figure}

Finally, we note that Eq.~\ref{eq:eq8dipole},
yields an effective macroscopic susceptibility scaling as
{
$\chi^H \propto p\,\Omega^{-q} \equiv \chi^H(q)$, with $q = 1/2$.
To assess whether the choice $q = 1/2$ optimizes the agreement with the continuous microscopic models, Fig.~\ref{fig:fig3} compares the previously introduced linear model with $\chi^H(q)$ evaluated for $q = 0$, $0.5$, and $1$ {, i.e. the original permittivity $\chi$}. Within the adopted approximations, the closest agreement is observed for $q = 1/2$.}

\begin{figure}
\includegraphics[scale=0.33]{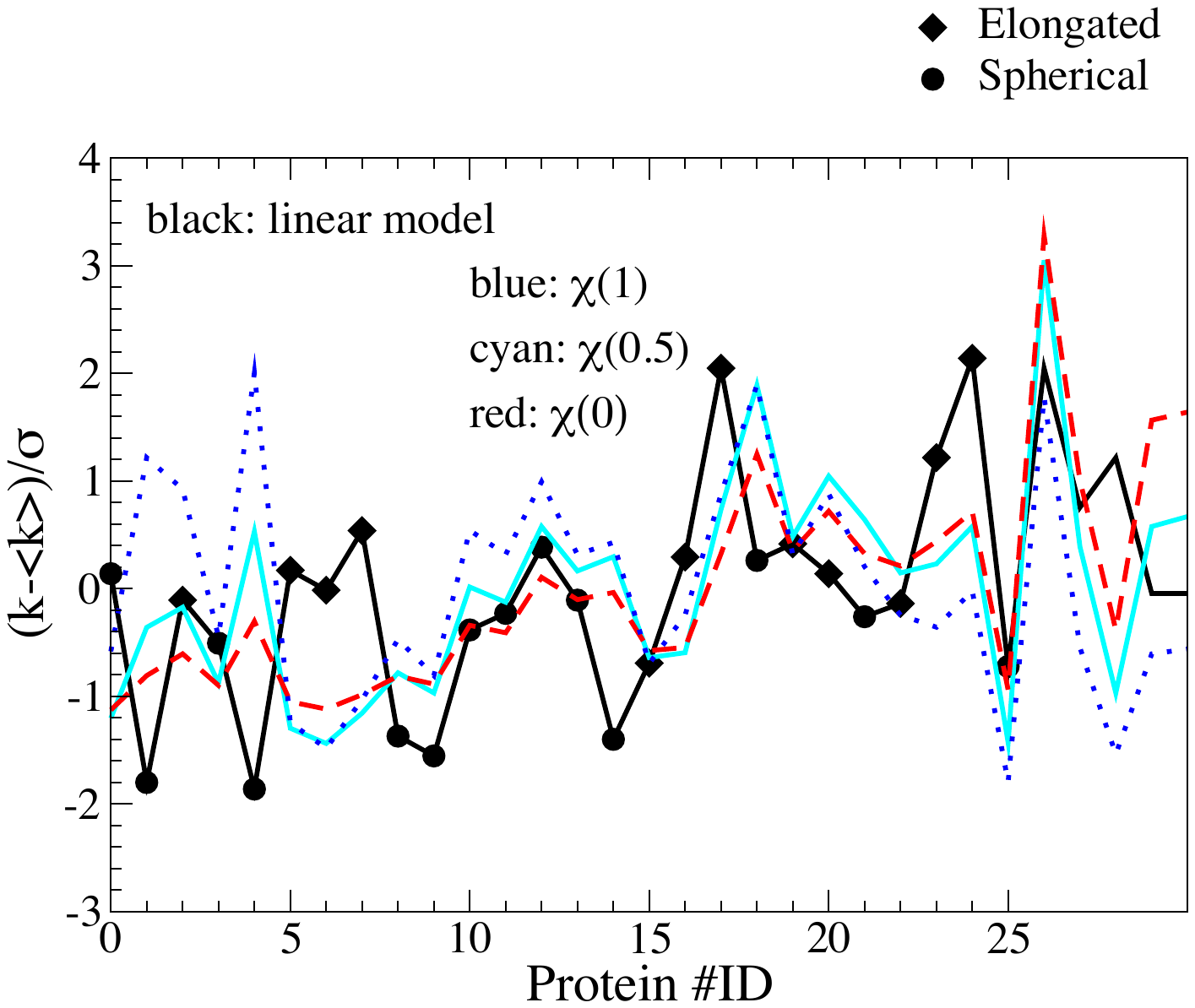}
\caption{\label{fig:fig3} 
Normalized \textit{macroscopic} susceptibility for the selected dataset: role of volume. The normalized permittivity values obtained using three different power-law scalings ($\chi \propto \Omega^{-n}$, with $n = 0$, $0.5$, and $1$) are compared with those calculated using the linear test function (black line); see also the main text.
}
\end{figure}

\medskip
\section{Discussions and conclusions}
The calculation and definition of permittivity in proteins remain subjects of ongoing debate and are known to depend sensitively on the adopted theoretical and computational framework (see, \textit{e.g.}, Ref.~\onlinecite{warshel2006modeling}).
{ On the other hand, this topic is of outmost importance for both fundamental and applicative researches, as for example, the development of biosensors or, in general, medical-oriented studies\cite{torculas2016protein,valagiannopoulos2021maximal,
huang2024capacitive,di2025electric}, therefore research is continuously evolving and exploring different strategies for interpreting and calculating the of main characteristics of the phenomenon \cite{warshel2006modeling,li2013dielectric,sauer2024linear,seyedi2018dipolar,amin2020variations,hartl2022dipolar,sauer2024linear, tenenbaum2024energy, agelii2025dipole}. }

In this paper, we propose a computationally efficient 
approach for estimating the effective macroscopic susceptibility of hydrated proteins. This formulation is motivated by the need to describe the orientational response of macromolecules with restricted rotational mobility arising from their interaction with a polar solvent.

{The investigation illustrated here is divided into two steps. In the first, we propose a microscopic study of the permittivity, $k$, of hydrated proteins, analyzing the interaction between individual amino acids and the solvent. This is done using a continuous distribution model of the permittivity of individual amino acids, assigning a higher value to those closest to the solvent molecules. This description simply translates the idea that while there is an \textit{intrinsic} (dry) permittivity of the amino acids, the same observed in the inner part of the protein, whose \textit{effective} value is provided by the simultaneous presence of the amino acids and the solvent. In electrical engineering, this is equivalent to a parallel circuit between the solvent and the biological material and is in perfect agreement with the description used in more refined procedures, such as, for example, } Ref.\onlinecite{simonson1996charge}.
{ The measured permittivity of a hydrated protein is derived from the effective dielectric constant. On the other hand, the permittivity of the same protein, in the absence of water, is simply derived from the intrinsic permittivity.}

The second step involves the macroscopic analysis of permittivity, {starting from the whole protein intrinsic dipole. In classical electrodynamics, in the presence of an intrinsic dipole, its orientation provides the main contribution to the shielding of electric fields: This response is quantified macroscopically in terms of the material's susceptibility and, ultimately, its permittivity. In particular,} a system of freely rotating dipoles yields the susceptibility given by Eq.~\ref{eq:eq3dipole}. This expression is not directly applicable to hydrated proteins (dipole moment $p$), as it leads to susceptibility values significantly larger than those reported experimentally (see Table~\ref{tab:table1}). The discrepancy arises from the fact that proteins in aqueous environments are not freely orientable: their rotational dynamics are constrained by solvent interactions {(\textit{i.e.}, viscosity)}, which in turn limit their dielectric response.
By contrast, the susceptibility evaluated using Eq.~\ref{eq:eq8dipole} produces values qualitatively consistent with the expected range \cite{simonson2003electrostatics}(see Table~\ref{tab:table1}). Furthermore, the susceptibility associated with dry proteins (dipole moment $p_0$), denoted as $\chi_0$, falls within the range typically attributed to the internal regions of proteins \cite{simonson2003electrostatics}.
We emphasize that the difference between $p$ and $p_0$ does not primarily originate from structural modifications of the three-dimensional conformation, but rather from changes in polarizability. The dipole moment $p$ characterizes a hydrated protein, where interfacial water molecules dynamically reorient and influence the distribution of surface charges, thereby altering the dielectric response, particularly for solvent-exposed residues. In contrast, $p_0$ corresponds to the limiting case of a dry protein, in which the native structure is ideally preserved while solvent-induced polarization and screening effects are absent.
Within this interpretation, $p$ represents a hydration-dependent state derived from $p_0$, where the dominant variation concerns the effective permittivity. Accordingly, $p$ may be regarded as a function of $p_0$, and its expectation value can be estimated using the orientational probability distribution associated with the freely rotating dipole $p_0$.
The results obtained using the two independent procedures show qualitative consistency, supporting the validity of the proposed framework. { The comparison of results from two different computational and design approaches is in itself a step forward in understanding the multiple aspects of permittivity in proteins, an issue which is well described by several authors and, in particular,  in Refs.\onlinecite{schutz2001dielectric,warshel2006modeling}. } Nevertheless, further refinements are possible, particularly regarding the selection of test functions, the choice of the interaction radius, $R_C$ and the evaluation of dipole moments.
Finally, we note that in physiological and experimental conditions proteins are generally not freely mobile. This observation suggests that additional investigation is warranted to clarify the definition and interpretation of electrical susceptibility in constrained biological environments.

{The proposed procedure is not intended to provide a complete or fully rigorous description. Rather, it offers a practical strategy that may complement more detailed molecular dynamics approaches, which, despite their accuracy, have not yet produced a comprehensive reference dataset of dielectric properties across diverse protein classes.}

\medskip
\centerline{AUTHOR DECLARATIONS}

\noindent\textbf{Conflict of Interests}\\
The Authors have no conflict of interest to disclose.\\
\textbf{Author Contribution}\\
E.A and M.B  contributed equally to this paper.\\
 \textbf{Data Availability}\\
 The data that support the findings of this paper are available from the corresponding author upon reasonable request.
\bibliography{TP_biblio}

@PREAMBLE{
 "\providecommand{\noopsort}[1]{}" 
 # "\providecommand{\singleletter}[1]{#1}%" 
}

@article{onsager1936electric,
  title={Electric moments of molecules in liquids},
  author={Onsager, Lars},
  journal={Journal of the American Chemical Society},
  volume={58},
  number={8},
  pages={1486--1493},
  year={1936},
  publisher={ACS Publications}
}

@article{kirkwood1936statistical,
  title={Statistical Mechanics of Liquid Solutions.},
  author={Kirkwood, John G},
  journal={Chemical Reviews},
  volume={19},
  number={3},
  pages={275--307},
  year={1936},
  publisher={ACS Publications}
}

@article{kirkwood1939dielectric,
  title={The dielectric polarization of polar liquids},
  author={Kirkwood, John G},
  journal={The Journal of Chemical Physics},
  volume={7},
  number={10},
  pages={911--919},
  year={1939},
  publisher={American Institute of Physics}
}

@misc{frolich1949theory,
  title={Theory of Dielectrics, Clarendon},
  author={Frolich, H},
  year={1949},
  publisher={Oxford}
}

@article{tanford1957location,
  title={The location of electrostatic charges in Kirkwood's model of organic ions},
  author={Tanford, Charles},
  journal={Journal of the American Chemical Society},
  volume={79},
  number={20},
  pages={5348--5352},
  year={1957},
  publisher={ACS Publications}
}

@article{tanford1957theory,
  title={Theory of protein titration curves. I. General equations for impenetrable spheres},
  author={Tanford, Charles and Kirkwood, John G},
  journal={Journal of the American Chemical Society},
  volume={79},
  number={20},
  pages={5333--5339},
  year={1957},
  publisher={ACS Publications}
}

@article{donnelly1994seven,
  title={Seven-helix receptors: structure and modelling},
  author={Donnelly, Dan and Findlay, John BC},
  journal={Current Opinion in Structural Biology},
  volume={4},
  number={4},
  pages={582--589},
  year={1994},
  publisher={Elsevier}
}

@article{antosiewicz1995computation,
  title={Computation of the dipole moments of proteins},
  author={Antosiewicz, Jan},
  journal={Biophysical journal},
  volume={69},
  number={4},
  pages={1344--1354},
  year={1995},
  publisher={Elsevier}
}

@article{harpaz1994volume,
  title={Volume changes on protein folding},
  author={Harpaz, Yehouda and Gerstein, Mark and Chothia, Cyrus},
  journal={Structure},
  volume={2},
  number={7},
  pages={641--649},
  year={1994},
  publisher={Elsevier}
}

@article{mo1995characteristics,
  title={Characteristics of dielectric behavior in nanostructured materials},
  author={Mo, Chi-mei and Zhang, Lide and Wang, Guozhong},
  journal={Nanostructured Materials},
  volume={6},
  number={5-8},
  pages={823--826},
  year={1995},
  publisher={Elsevier}
}

@article{simonson1996charge,
  title={Charge screening and the dielectric constant of proteins: insights from molecular dynamics},
  author={Simonson, Thomas and Brooks, Charles L},
  journal={Journal of the American Chemical Society},
  volume={118},
  number={35},
  pages={8452--8458},
  year={1996},
  publisher={ACS Publications}
}

@article{vendruscolo1997recovery,
  title={Recovery of protein structure from contact maps},
  author={Vendruscolo, Michele and Kussell, Edo and Domany, Eytan},
  journal={Folding and Design},
  volume={2},
  number={5},
  pages={295--306},
  year={1997},
  publisher={Elsevier}
}

@article{warshel1998,
  title={The effect of protein relaxation on charge-charge interactions and dielectric constants of proteins},
  author={Sham, Yuk Yin and Muegge, Ingo and Warshel, Arieh},
  journal={Biophysical Journal},
  volume={74},
  number={4},
  pages={1744--1753},
  year={1998},
  publisher={Elsevier}
}

@article{havranek1999tanford,
  title={Tanford--Kirkwood electrostatics for protein modeling},
  author={Havranek, James J and Harbury, Pehr B},
  journal={Proceedings of the National Academy of Sciences},
  volume={96},
  number={20},
  pages={11145--11150},
  year={1999},
  publisher={The National Academy of Sciences}
}

@article{berman2000protein,
  title={The protein data bank},
  author={Berman, Helen M and Westbrook, John and Feng, Zukang and Gilliland, Gary and Bhat, Talapady N and Weissig, Helge and Shindyalov, Ilya N and Bourne, Philip E},
  journal={Nucleic acids research},
  volume={28},
  number={1},
  pages={235--242},
  year={2000},
  publisher={Oxford University Press}
}

@article{schutz2001dielectric,
  title={What are the dielectric “constants” of proteins and how to validate electrostatic models?},
  author={Schutz, Claudia N and Warshel, Arieh},
  journal={Proteins: Structure, Function, and Bioinformatics},
  volume={44},
  number={4},
  pages={400--417},
  year={2001},
  publisher={Wiley Online Library}
}

@article{grant2001smooth,
  title={A smooth permittivity function for Poisson--Boltzmann solvation methods},
  author={Grant, J Andrew and Pickup, Barry T and Nicholls, Anthony},
  journal={Journal of computational chemistry},
  volume={22},
  number={6},
  pages={608--640},
  year={2001},
  publisher={Wiley Online Library}
}

@article{de2000blue,
  title={Blue copper proteins: a comparative analysis of their molecular interaction properties},
  author={DE RIENZO, Francesca and Gabdoulline, RR and Menziani, Maria Cristina and Wade, RC},
  journal={Protein Science},
  volume={9},
  number={8},
  pages={1439--1454},
  year={2000},
  publisher={Cambridge University Press}
}

@article{song2002inhomogeneous,
  title={An inhomogeneous model of protein dielectric properties: Intrinsic polarizabilities of amino acids},
  author={Song, Xueyu},
  journal={The Journal of chemical physics},
  volume={116},
  number={21},
  pages={9359--9363},
  year={2002},
  publisher={American Institute of Physics}
}

@article{simonson2003electrostatics,
  title={Electrostatics and dynamics of proteins},
  author={Simonson, Thomas},
  journal={Reports on Progress in Physics},
  volume={66},
  number={5},
  pages={737},
  year={2003},
  publisher={IOP Publishing}
}

@book{riande2004electrical,
  title={Electrical properties of polymers},
  author={Riande, Evaristo and D{\'\i}az-Calleja, Ricardo},
  year={2004},
  publisher={CRC Press}
}

@article{warshel2006modeling,
  title={Modeling electrostatic effects in proteins},
  author={Warshel, Arieh and Sharma, Pankaz K and Kato, Mitsunori and Parson, William W},
  journal={Biochimica et Biophysica Acta (BBA)-Proteins and Proteomics},
  volume={1764},
  number={11},
  pages={1647--1676},
  year={2006},
  publisher={Elsevier}
}

@article{felder2007server,
  title={A server and database for dipole moments of proteins},
  author={Felder, Clifford E and Prilusky, Jaime and Silman, Israel and Sussman, Joel L},
  journal={Nucleic acids research},
  volume={35},
  number={suppl\_2},
  pages={W512--W521},
  year={2007},
  publisher={Oxford University Press}
}

@book{cardarelli2008materials,
  title={Materials handbook: a concise desktop reference},
  author={Cardarelli, Fran{\c{c}}ois},
  year={2008},
  publisher={Springer}
}

@MISC{molecularmass, 

   author       = "",
   title        = "2020-09-17 Jimmy Eng. UW Proteomics Resource", 
   howpublished = "http://www.umimod.org/masses.html", 
   month        = "09", 
   year         = "2020", 
   note         = "",
}

@article{alfinito2008network,
  title={A network model to correlate conformational change and the impedance spectrum of singleproteins},
  author={Alfinito, Eleonora and Pennetta, Cecilia and Reggiani, Lino},
  journal={Nanotechnology},
  volume={19},
  number={6},
  pages={065202},
  year={2008},
  publisher={IOP Publishing}
}

@article{simonson2008dielectric,
  title={Dielectric relaxation in proteins: the computational perspective},
  author={Simonson, Thomas},
  journal={Photosynthesis research},
  volume={97},
  number={1},
  pages={21--32},
  year={2008},
  publisher={Springer}
}

@article{alfinito2010single,
  title={A single protein based nanobiosensor for odorant recognition},
  author={Alfinito, Eleonora and Millithaler, J-F and Pennetta, Cecilia and Reggiani, Lino},
  journal={Microelectronics journal},
  volume={41},
  number={11},
  pages={718--722},
  year={2010},
  publisher={Elsevier}
}

@article{alfinito2011human,
  title={Human olfactory receptor 17-40 as an active part of a nanobiosensor: a microscopic investigation of its electrical properties},
  author={Alfinito, Eleonora and Millithaler, Jean-Francois and Reggiani, Lino and Zine, Nadia and Jaffrezic-Renault, Nicole},
  journal={Rsc Advances},
  volume={1},
  number={1},
  pages={123--127},
  year={2011},
  publisher={Royal Society of Chemistry}
}

@article{matyushov2012dipolar,
  title={Dipolar response of hydrated proteins},
  author={Matyushov, Dmitry V},
  journal={The Journal of chemical physics},
  volume={136},
  number={8},
  year={2012},
  publisher={AIP Publishing}
}

@article{li2013dielectric,
  title={On the dielectric “constant” of proteins: smooth dielectric function for macromolecular modeling and its implementation in DelPhi},
  author={Li, Lin and Li, Chuan and Zhang, Zhe and Alexov, Emil},
  journal={Journal of chemical theory and computation},
  volume={9},
  number={4},
  pages={2126--2136},
  year={2013},
  publisher={ACS Publications}
}

@inproceedings{eleonora2015proteotronics,
  title={Proteotronics: Electronic devices based on proteins},
  author={Alfinito, Eleonora and Reggiani, Lino and Pousset, Jeremy},
  booktitle={Sensors: Proceedings of the Second National Conference on Sensors, Rome 19-21 February, 2014},
  pages={3--7},
  year={2015},
  organization={Springer}
}

@article{chen2015proteinvolume,
  title={ProteinVolume: calculating molecular van der Waals and void volumes in proteins},
  author={Chen, Calvin R and Makhatadze, George I},
  journal={BMC bioinformatics},
  volume={16},
  number={1},
  pages={101},
  year={2015},
  publisher={Springer}
}

@article{bellissent2016water,
  title={Water determines the structure and dynamics of proteins},
  author={Bellissent-Funel, Marie-Claire and Hassanali, Ali and Havenith, Martina and Henchman, Richard and Pohl, Peter and Sterpone, Fabio and Van Der Spoel, David and Xu, Yao and Garcia, Angel E},
  journal={Chemical reviews},
  volume={116},
  number={13},
  pages={7673--7697},
  year={2016},
  publisher={ACS Publications}
}

@article{torculas2016protein,
  title={Protein-based bioelectronics},
  author={Torculas, Maria and Medina, Jethro and Xue, Wei and Hu, Xiao},
  journal={ACS Biomaterials Science \& Engineering},
  volume={2},
  number={8},
  pages={1211--1223},
  year={2016},
  publisher={ACS Publications}
}

@article{seyedi2018dipolar,
  title={Dipolar susceptibility of protein hydration shells},
  author={Seyedi, Salman and Matyushov, Dmitry V},
  journal={Chemical Physics Letters},
  volume={713},
  pages={210--214},
  year={2018},
  publisher={Elsevier}
}

@article{amin2020variations,
  title={Variations in proteins dielectric constants},
  author={Amin, Muhamed and K{\"u}pper, Jochen},
  journal={ChemistryOpen},
  volume={9},
  number={6},
  pages={691--694},
  year={2020},
  publisher={Wiley Online Library}
}

@article{valagiannopoulos2021maximal,
  title={Maximal interaction of electromagnetic radiation with corona virions},
  author={Valagiannopoulos, Constantinos and Sihvola, Ari},
  journal={Physical Review B},
  volume={103},
  number={1},
  pages={014114},
  year={2021},
  publisher={APS}
}

@article{hartl2022dipolar,
  title={Dipolar interactions and protein hydration in highly concentrated antibody formulations},
  author={Hartl, Josef and Friesen, Sergej and Johannsmann, Diethelm and Buchner, Richard and Hinderberger, Dariush and Blech, Michaela and Garidel, Patrick},
  journal={Molecular Pharmaceutics},
  volume={19},
  number={2},
  pages={494--507},
  year={2022},
  publisher={ACS Publications}
}

@article{sauer2024linear,
  title={Linear and nonlinear dielectric response of intrinsically disordered proteins},
  author={Sauer, Michael A and Colburn, Taylor and Maiti, Sthitadhi and Heyden, Matthias and Matyushov, Dmitry V},
  journal={The Journal of Physical Chemistry Letters},
  volume={15},
  number={20},
  pages={5420--5427},
  year={2024},
  publisher={ACS Publications}
}

@article{huang2024capacitive,
  title={Capacitive biosensors for label-free and ultrasensitive detection of biomarkers},
  author={Huang, Lei and Zhang, Cheng and Ye, Run and Yan, Bin and Zhou, Xiaojia and Xu, Wenbo and Guo, Jinhong},
  journal={Talanta},
  volume={266},
  pages={124951},
  year={2024},
  publisher={Elsevier}
}

@article{tenenbaum2024energy,
  title={Energy condensation and dipole alignment in protein dynamics},
  author={Tenenbaum, Alexander},
  journal={Physical Review E},
  volume={109},
  number={4},
  pages={044401},
  year={2024},
  publisher={APS}
}

@article{agelii2025dipole,
  title={Dipole orientation of hydrated gas phase proteins},
  author={Agelii, Harald and Jakobsson, Ellen LS and De Santis, Emiliano and Elfrink, Gideon and Mandl, Thomas and Marklund, Erik G and Caleman, Carl},
  journal={Physical Chemistry Chemical Physics},
  volume={27},
  number={21},
  pages={10939--10948},
  year={2025},
  publisher={Royal Society of Chemistry}
}

@article{di2025electric,
  title={Electric field cycling of physisorbed antibodies reduces biolayer polarization dispersion},
  author={Di Franco, Cinzia and Macchia, Eleonora and Catacchio, Michele and Caputo, Mariapia and Scandurra, Cecilia and Sarcina, Lucia and Bollella, Paolo and Tricase, Angelo and Innocenti, Massimo and Funari, Riccardo and others},
  journal={Advanced Science},
  volume={12},
  number={1},
  pages={2412347},
  year={2025},
  publisher={Wiley Online Library}
}


\appendix
\section{\label{app:appA}Model function}

Within the proteotronics framework, each protein conformation is represented as an interaction network in which nodes correspond to amino acids centroids \cite{alfinito2008network,alfinito2011human,alfinito2010single,alfinito2008network}. Links between nodes are introduced when the inter-residue distance is smaller than a prescribed interaction radius, $R_C$, taken here as 6 \AA. Under this construction, each node is characterized by a coordination number, i.e., the number of nearest neighbors, which depends on its position within the network.
Model distribution functions are employed to assign a local dielectric constant, $\epsilon(n)$, to the $n$th amino acid. This parameter varies between the intrinsic value $\epsilon_i(n)$ \cite{song2002inhomogeneous} and $\mu$, representing the dielectric permittivity of the solvent.

\vskip 0.2cm
\noindent\textbf{Model 1:} polynomial
\begin{equation}
    \epsilon(n)=\epsilon_i(n)\,x^m(n)+\mu\, (1-x^m(n)),
\end{equation}

\vskip 0.2cm
\noindent\textbf{Model 2:} sigmoidal
\begin{eqnarray}
    &\epsilon(n) &=\epsilon_i(n)\,x(n)\,w+\mu\,(1-x(n))\,(1-w), \nonumber \\
    &w &= \frac{L^q(n)}{L^q(n)+\langle L\rangle^q}.
\end{eqnarray}

\vskip 0.2cm
\noindent\textbf{Model 3:} exponential
\begin{equation}
    \epsilon(n)=\epsilon_i(n)\,x(n)+\mu\, (1-x(n))\, 
    \exp\left[-\left(\frac{L(n)}{\langle L\rangle}\right)^q\right],
\end{equation}
with $x(n) = \frac{L(n)}{L_{\text{max}}}$, where $L(n)$ denotes the coordination number associated with the $n$th amino acid, $L_{\text{max}}$ is the maximum coordination number observed among all amino acids in the protein and $\langle L\rangle$ is the average value.

In all models, the parameters $q$ and $m$ control the sharpness of the  function and, consequently, the spatial extent of the region where $\epsilon(n)$ assumes its largest values. From a physical standpoint, these parameters effectively describe the degree of protein hydration.
In the present work, the three models are implemented using two distinct parameter sets, specifically:

\medskip\noindent
Model 1: $m=1,0.5$ (linear and polynomial);

\medskip\noindent
Model 2: $q=1,6$ (mild and sharp sigmoidal)

\medskip\noindent
Model 3: $q=1,6$ (mild and sharp exponential)

\section{\label{app:appB}Tables}

Table~\ref{tab:table1} summarizes, for the analyzed protein dataset, the dipole moment $p$ obtained from Ref.~\onlinecite{felder2007server}, the effective susceptibilities of hydrated ($\chi^H$) and dry ($\chi_0$) proteins, and the dipole moment of the dry protein, $p_0$, as estimated in this {paper}. Dipole moments are expressed in debye (D).
 
The intrinsic permittivity of each amino acid, $\epsilon_i$, is evaluated using the Clausius–Mossotti relation together with the molecular polarizabilities, $\alpha$, reported in Ref.~\onlinecite{song2002inhomogeneous}. Specifically,
\begin{equation}
\frac{4 n \pi \alpha}{3} = \frac{\epsilon_i - 1}{\epsilon_i + 2},
\label{eq:clausius}
\end{equation}
where $n = \rho \mathcal{N}_A / A$, with $\mathcal{N}_A$ denoting Avogadro’s number, $\rho$ the molecular density \cite{cardarelli2008materials}, and $A$ the mass number \cite{molecularmass}.

{The list of intrinsic permittivities is resumed} in Table~\ref{tab:table2}.

{ 
Finally,the dipole of the dry protein, $p_0$, has been calculated by using $\chi_0=k_0-1$, where $k_0$ represents the mean value of the intrinsic permittivities, $\epsilon_i$ of the amino acids present in the protein : 
\begin{equation}
    {p_0}^2=3\epsilon_0 k_BT\Omega \chi_0
\end{equation}
 which, by using for the temperature the standard value $T=300K$, and expressing $p_0$ in debyes, gives: ${p_0}^2 \approx 1.26\chi_0 N_{aa}. $}

\begin{table*}
\caption{\label{tab:table1} List  of proteins used in this paper. Dipole values are given in debyes. Notice that for complex proteins like 1p9m, 2gy7,2j8c  both  whole structure (biological unit) and  single chains -(protein$_{chain}$)- are reported.  }
\begin{ruledtabular}
\begin{tabular}{llccccc}
ID & \footnote{Ref.~\onlinecite{berman2000protein}}Code & $N_{aa}$& \footnote{Ref.~\onlinecite{felder2007server}}$p$(D)&\footnote{using $T$=300K}$\chi^H$ &$p_0$(D) & $\chi_0$\\
\hline

0	&1c09	&53&	74&	6	&	6 &0.48\\
1	&1c9o &	66	& 200	&15	&	6&0.65 \\
2	& 1uz3 &	102	&280&	17&	8&0.56 \\
3	&1brk &	108 &	164	&10	&	8&0.62\\
4	&1bkr&108 &	400	&	24& 8 &0.62\\
5& 	1nzr&	128&	104	&	6& 9&0.65 \\
6& 1aiz	&129 &	77	& 4 &		9&0.59\\
7& 1e65 &	129 &	130& 		7 & 9&0.66 \\
8&	3lzt &	129	&198&	11	&9&0.60\\
9 &	1b0b	&141&	169&	9&	9&0.55\\
10 &1alu&	157&	385	&	19& 10&0.72\\
11&	1p9m$_B$&	163&	358&		17&	10&0.72\\
12&	1p9m$_C$&	201 &	562	&25 &	11&0.70\\
13&	1z3s	&216&	480	&20&12&0.65\\
14& 2gy7$_A$&	216&	506&	21& 12&0.65\\
15&	2ntw	&222&	292&	12& 12&0.37\\
16 &	2ntu	&222	&304	&13 &12&0.37\\	
17& 2j8c$_H$	&241	&649	&26&	12&0.49\\
18& 2j8c$_L$&	281	&1015&		38& 13&0.42\\
19&	1p9m$_A$&	298&	660&		24  &14&0.63\\
20& 1n26&	299&	805& 	29 & 14&0.70\\
21& 2j8c$_M$ &	302&	649	& 	23& 14&0.47\\
22&	1u19& 	348 &	604& 20& 15&0.50 \\	
23& 2gy7$_B$ &	423&	694& 21& 16&0.55\\	
24&	2gy5&	423&	804& 24&16&0.55\\	
25&	3ehs	& 457&	148& 4&17&0.55 \\	
26&	2ace&	527 &	1819& 49&18&0.56\\	
27&	2gy7&	639	&919&23&20&0.58\\	
28&	1p9m&662&	368 &9 &20&0.58 \\	
29 &	2j8c&	825&	1140&	25& 23&0.46\\
30 &	2uxj	&825&	1170	&25& 23&0.46 \\

\end{tabular}
\end{ruledtabular}
\end{table*}

\begin{table*}
\caption{\label{tab:table2}List of main data used for the calculation of intrinsic dielectric constant $\epsilon_i$}
\begin{ruledtabular}
    \begin{tabular}{lcclcc}
    amino acid&	 \footnote{mass number: http://www.unimod.org/masses.html, 2025-09-24 Jimmy Eng, UW Proteomics Resource} $A$ &\footnote{
    https://matmake.com/properties/density-of-amino-acids.html}	$\rho$(g/\AA$^3$)&	\footnote{Ref.~\onlinecite{cardarelli2008materials}}n(mol/\AA$^3$)	& \footnote{Ref.~\onlinecite{song2002inhomogeneous} }$\alpha$ (\AA$^3$)&$\epsilon_i$ 
    \\ \\
    \hline
alanine	& 71 &	1.430&	1.12 $\times 10^{-2}$ &	1.1&		1.18 \\	
arginine &156 &	1.230&	0.47$\times 10^{-2}$ &	5.7&		1.38 \\	
asparagine	&	114	&1.543&	0.82$\times 10^{-2}$& 9.8	&2.51\\	
aspartic acid	&	115	&1.700&	0.89$\times 10^{-2}$ &	4.0&	1.52\\	
cysteine	& 103	& 1.180 &	0.69$\times 10^{-2}$ & 	1.8&	1.16	\\
glutamic acid	& 129	& 1.538 & 0.72$\times 10^{-2}$ & 6.2&	1.69\\
glutamine	& 128	& 1.364  & 	0.64$\times 10^{-2}$ & 17.7	& 3.72\\	
glycine	& 	57 & 	1.161 & 	1.23$\times 10^{-2}$ & 	2.2& 	1.38\\	
histidine	&	137	& 1.309  &	0.57$\times 10^{-2}$ & 	12.3&2.26\\	
isoleucine	&	113&	1.293 &	0.69$\times 10^{-2}$ &	1.1& 	1.10\\	
leucine	&	113& 	1.293  &	0.69$\times 10^{-2}$ & 	1.2 &	1.11	\\
lysine	& 128 &	1.136  &	0.53$\times 10^{-2}$ &	8.4&	1.70	\\
methionine	&	131	&1.178  &	0.54$\times 10^{-2}$ &	2.8 &1.20	\\
phenylalanine	& 147& 	1.160 &	0.47$\times 10^{-2}$& 1.0& 1.06\\	
proline	& 97 & 	1.180& 	0.73	$\times 10^{-2}$ & 	1.3& 	1.12\\
serine	&	87& 	1.600& 	1.11$\times 10^{-2}$ &	7.3&		2.54	\\
threonine	& 101&	1.131 &	0.67$\times 10^{-2}$& 8.2&	1.90	\\
tryptophan	& 186& 	1.175 &	0.38$\times 10^{-2}$&	1.8&	1.09	\\
tyrosine	& 163 &	1.237 &	0.46$\times 10^{-2}$&	5.3&	1.34	\\
valine	&99&	1.230&	0.75$\times 10^{-2}$&	1.0&	1.10	
    \end{tabular}

\end{ruledtabular}
\end{table*}

\begin{acknowledgments}
We thank Professors Joel Sussman and Jaime Prilusky (Weizmann Institute of Science, Rehovot, Israel) for helpful discussions regarding the Protein Dipole Moment Server.

This paper has been partially funded by the Italian MUR project GINEVRA, prot. 2022BZYBWM.
\end{acknowledgments}
\end{document}